\documentclass[
  aps,
  prb,
  reprint,
  amssymb,
  superscriptaddress,
  longbibliography
]{revtex4-2}
\usepackage[dvipsnames]{xcolor}
\usepackage[colorlinks=true,
            linkcolor=blue,
            citecolor=blue,
            urlcolor=blue]{hyperref}
\usepackage{physics} 
\usepackage{color}
\usepackage{subfig}
\usepackage{subcaption}
\usepackage{graphicx}

\begin{document}

%Title of paper
\title{Phonon-assisted charge-cycling of nitrogen-vacancy centres in diamond}%

\author{Michael Olney-Fraser}%
\email[Corresponding author: ]{michael.olney-fraser@uni-ulm.de}
\affiliation{Institute for Quantum Optics, Ulm Univesity, Albert-Einstein-Allee 11, 89081 Ulm, Germany}
\affiliation{Integrated Quantum Science and Technology (IQST), Ulm University, Albert-Einstein-Allee 11, Ulm 89081, Germany}

\author{Stefan Dietel}
\affiliation{Institute for Quantum Optics, Ulm Univesity, Albert-Einstein-Allee 11, 89081 Ulm, Germany}

\author{Jens Fuhrmann}
\affiliation{Institute for Quantum Optics, Ulm Univesity, Albert-Einstein-Allee 11, 89081 Ulm, Germany}

\author{Lev Kazak}
\affiliation{Institute for Quantum Optics, Ulm Univesity, Albert-Einstein-Allee 11, 89081 Ulm, Germany}

\author{Fedor Jelezko}%
\affiliation{Institute for Quantum Optics, Ulm Univesity, Albert-Einstein-Allee 11, 89081 Ulm, Germany}
\affiliation{Integrated Quantum Science and Technology (IQST), Ulm University, Albert-Einstein-Allee 11, Ulm 89081, Germany}

\date{\today}

\begin{abstract}

The nitrogen-vacancy (NV) centre in diamond is a leading platform for room-temperature quantum sensing. Improvements in sensitivity require precise control of the NV charge state. Transitions from the neutral NV$^0$ charge state to the negative NV$^-$ charge state can occur during excitation with photon energies below the ZPL transition of NV$^0$. These sub-resonant charge transitions limit modern initialisation protocols and have not been studied in full detail. In this paper we show that sub-resonant charge cycling arises from phonon-assisted anti-Stokes excitation. We further uncover the phonon states which contribute most strongly to the anti-Stokes transition via the development of novel quantitative models. The models indicate that low energy acoustic phonons strongly contribute to the transition close to the ZPL. At longer wavelengths a 43\,meV mode additionally impacts the charge dynamics.

\end{abstract}

\maketitle

\section{Introduction}

The nitrogen-vacancy (NV) centre in diamond is a leading platform for quantum sensing and metrology owing to its nanoscale sensitivity to magnetic and electric fields at room temperature \cite{aslam23, Acosta09, julia19}. The sensitivity of NV-based quantum sensors depends strongly on the fidelity of optical spin initialisation and readout. However, these fidelities are limited by unwanted transitions between the two primary NV charge states \cite{barry20}.

The negative charge state, NV$^-$, has a spin-1 electronic ground state with long coherence times at room temperature \cite{bala09}. For this reason, NV$^-$ is the charge state most commonly used for sensing applications. Conversely the neutral charge state, NV$^0$, has received less attention, although recent work has begun to uncover its potentially useful properties \cite{gali09, barson19, baier20}. Understanding and controlling transitions between these states is key to high fidelity optical initialisation and readout. 

One such initialisation and readout technique involves exciting the system with photons whose energies sit between the zero-phonon lines (ZPL) of NV$^-$ (637\,nm) and NV$^0$ (575\,nm) \cite{sheilds2015, jaskula19, hopper2020}. In this regime, the transition from NV$^0$ to NV$^-$ is suppressed as the excitation photons do not provide enough energy to excite NV$^0$ directly. However, this transition is still routinely observed despite being energetically forbidden \cite{wood24, aslam13, gao22}. The primary interpretation of this behaviour is that the excitation of NV$^0$ is facilitated by the simultaneous absorption of a photon and one or more phonons, where the phonons provide the extra energy required to drive the transition, this is known as anti-Stokes shifted excitation \cite{wood24, gao22}. Existing studies of this transition have, however, been performed on NV ensembles, where the phononic states responsible for the process cannot be resolved  \cite{wood24, gao22}. Additionally, separate charge-transfer mechanisms such as electron tunnelling and the ionisation of external defects may also contribute to the dynamics in NV ensembles \cite{Giri19}. As a result, the microscopic origin of the transition and the vibrational states involved have not been fully determined. We address this knowledge gap by determining whether phonon-assisted excitation alone can account for the observed charge dynamics and the vibrational structure responsible for the transition.

The present work focuses on charge transitions using a single, spatially isolated NV centre in high-purity diamond. The transition rate from NV$^0$ to NV$^-$ is extracted from photoluminescence blinking between the bright NV$^-$ and dark NV$^0$ charge states. We measure the transition rate across a range of excitation wavelengths and temperatures in order to probe the underlying mechanism behind the charge dynamics. The observed transition from NV$^0$ to NV$^-$ is shown to arise from phonon-assisted anti-Stokes excitation, evidenced by a strong suppression of the transition at cryogenic temperatures and a pronounced dependence on excitation photon energy detuning below the NV$^0$ ZPL. 

Direct insight into the vibrational environment which couples to the NV$^0$ electronic transition is gained via the development of quantitative models of the transition rate.

First, we introduce a simplified \textit{quasi-continuum model} in which the vibrational environment is derived from the NV$^0$ emission spectrum. Under several simplifying assumptions, this model captures the essential qualitative behaviour of the transition rate, especially at higher temperatures, and identifies the importance of low energy acoustic phonon modes in facilitating the transition. However, it is limited in its ability to resolve the individual vibrational modes which dominate the low temperature dependence of the transition rate.  

To overcome this limitation, we introduce a microscopic \textit{effective mode model}, in which a small number of modes are treated explicitly as harmonic oscillators and the transition rate is derived from Franck–Condon overlap integrals \cite{lax52}. Within this framework, we identify the interplay between low-energy acoustic phonons and a 43\,meV quasi-local vibrational resonance as the dominant contribution governing the phonon-assisted excitation process.

Together, the measurements and models provide a precise physical picture of anti-Stokes charge cycling in the NV centre. We demonstrate that sub-ZPL charge cycling in single NV centres is quantitatively explained by phonon-assisted anti-Stokes excitation and identify the dominant vibrational modes responsible. These results provide a framework for understanding sub-resonant charge initialization and readout techniques \cite{sheilds2015, jaskula19, hopper2020}, which offer improved readout fidelity and are important for high-sensitivity NV-based quantum sensors \cite{barry20}.

\begin{figure*}
    \centering
    \includegraphics[width=0.90\linewidth]{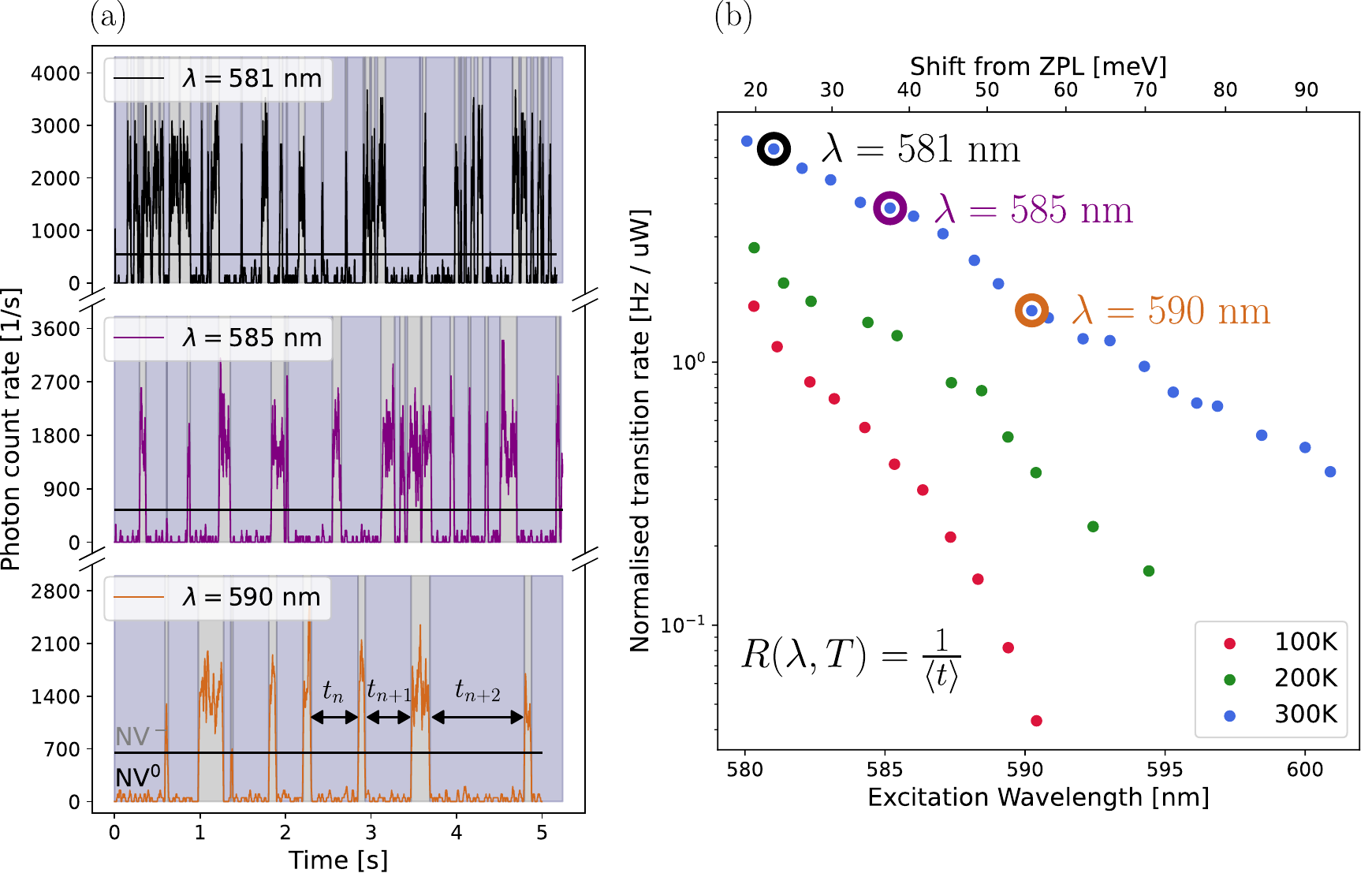}
    \caption{(a) Example photon count time series measured at 300\,K with excitation wavelengths of 581\,nm (black), 585\,nm (purple) and 590\,nm (orange). The blue shaded regions, when counts are below the threshold line, are attributed to NV$^0$. The NV$^-$ bright state is represented by the grey shaded areas. Dwell times $t_n,t_{n+1},t_{n+2}$ are shown in the bottom most plot. The averages of the NV$^0$ dwell time sets were inverted to find the transition rates, $R(\lambda, T)$ for each wavelength and temperature. (b) Transition rates as a function of excitation wavelength at temperatures 100\,K (red), 200\,K (green) and 300\,K (blue). The three wavelengths featured in plot (a) are indicated by correspondingly coloured circles.}
    \label{fig:timeseries}
\end{figure*}

\section{Methods}
\subsection{Experimental methods}

We used a type IIa, electronic-grade, single crystal, (100)-oriented diamond from ElementSix. In order to create intrinsic NV centres the diamond was annealed at 1500\,$^{\mathrm{\circ}}C$ for 3 hours in high vacuum. After annealing, solid immersion lenses (SILs) were fabricated on the diamond surface by focused ion beam milling. The SILs had a 5\,$\mu$m diameter and were designed for an objective with a numerical aperture of 0.95. The SILs enhanced the photoluminescent intensity of the NV used in this study significantly compared with NVs measured in bulk diamond. This enhancement enabled the photoluminescence signals in the bright and dark states to be accurately distinguished. 

The sample was mounted inside a continuous helium flow cryostat which could control the temperature in the range from $T$ = 4\,K to 300\,K. The NV was irradiated using a continuous wave, tunable laser and the measurements were conducted with excitation wavelengths from $\lambda$ = 580\,nm to 600\,nm and powers between 1\,$\mu$W and 5\,$\mu$W. 

The NV centre was imaged with a homebuilt confocal microscope and photon emissions were detected using a single-photon avalanche photo-diode (APD) filtered with a 650\,nm longpass filter. At each wavelength and temperature ($\lambda$, $T$), a long time series (approximately 500\,s) of photon emissions from the NV centre was measured. See Section I of the Supplemental Material (SM) \cite{SuppMat} for more information about the NV centre (and references \cite{brown56, zwiller01} therein).

\subsection{Transition rate measurement scheme}

The transition rate from NV$^0$ to NV$^-$ was measured by analysing the blinking statistics of the NV photoluminescence signal.

In Fig. \ref{fig:timeseries}(a), example time series data measured at 300\,K is shown. The bright state was attributed to the NV$^-$ charge state, where direct excitation of the main transition leads to a burst of photon emissions. After many cycles of excitation and emission, a photon is absorbed in the excited state and the system is ionised, returning it to NV$^0$. The dark state was attributed to the NV$^0$ charge state, where emissions are suppressed as the incident photons do not have the energy required to excite the system directly. This allowed a threshold value on the photon counts to be defined to discriminate between the charge states.

By recording a long time series for each excitation wavelength and temperature, large sets of ``dwell times" were gathered, defined as the length of time the system spends in one of the charge states before transitioning. Of interest to this investigation were the dwell times in the NV$^0$ dark state, which corresponded to the time taken for the system to transition from NV$^0$ to NV$^-$. With a large enough set, the mean value of the dwell times, $\langle t\rangle$, can be considered a statistically robust measure of the time the system spends in NV$^0$ on average. This mean dwell time could then be inverted to give the transition rate from NV$^0$ to NV$^-$, $R(\lambda, T)=1/\langle t\rangle$.  

\section{Experimental results}

Measurements were first performed at 4\,K. At this temperature, all the excited vibrational states are almost completely depopulated. We observed that the transition was strongly suppressed with rates which were too slow to be measured within the experimental limits (see Section II of the SM \cite{SuppMat} for the experimental data). This observation is informative as it indicates that parallel processes such as charge tunnelling or recharging from free charge carriers do not play a role in the transition for well isolated, single NV centres. These processes are not expected to depend on temperature as they are not facilitated by phonons. Thus, the suppression of the transition rate indicates that phonons are the essential mediator of the transition because their absence completely quenches the process.

At higher temperatures, statistically robust transition rates could be extracted from the time-series data. The full sets of transition rates measured at 100\,K, 200\,K and 300\,K across the wavelength range from 580\,nm to 600\,nm are shown in Fig. \ref{fig:timeseries}(b). These results show that as the excitation wavelength is red-shifted further from the NV$^0$ ZPL, the transition rate drops rapidly. Moreover, as the temperature increased, the transition rate increased. The strong temperature and wavelength dependence of the transition rate is another signature of a phonon-mediated process.

Further, the structure of the data encodes the vibrational environment which underpins the transition. In order to uncover this underlying structure, we develop quantitative models for the transition rate as a function of wavelength and temperature. 

\section{Transition rate modelling}\label{sec:model}

\subsection{General expression for the transition rate}

The transition from the ground state of NV$^0$ to the ground state of NV$^-$ consists of two transitions. Firstly, the system must be promoted to the excited state of NV$^0$. In the case of anti-Stokes shifted excitation, this transition requires the thermal occupation of vibrational states to provide the energy which is missing from the incident photons and, as a result, the transition rate is slow (Hz). Secondly, a photon must excite an electron from the valence band of the diamond into a bound state of the NV centre, recombining the system to NV$^-$. This transition can be driven directly by the incident photons leading to a much faster rate (MHz). The total transition rate from the ground state of NV$^0$ to NV$^-$ is therefore dominated by the slower transition between the ground and excited states of NV$^0$ (see Section III of the SM \cite{SuppMat} for a detailed derivation and reference \cite{allen10} therein). 

We can therefore express the total transition rate under optical excitation with wavelength $\lambda$ and photon flux density $\Phi_\lambda = P_\lambda / E_\lambda$, where $E_\lambda<E_{ZPL}^0$ is the photon energy and $P_\lambda$ is the laser power density as

\begin{equation}\label{eq:general_rate}
R(\lambda,T) = C\sigma(\lambda,T)\Phi_\lambda
\end{equation}

where $\sigma(\lambda,T)$ is the cross section of anti-Stokes shifted excitation from the ground state to the excited state of NV$^0$ and $C$ is a proportionality constant. 

The cross section of the anti-Stokes shifted transition is found by summing over the wavefunction overlaps between all the possible initial and final vibrational states, scaled by the thermal occupation of the initial states, and can be expressed as \cite{lax52}

\begin{align}
\sigma(\lambda, T) &= C^\prime E_\lambda|\hat{\mu}|^2\sum_{\{n_e\}}\sum_{\{n_g\}}\mathcal{B}(n_g,T)|\langle n_e|n_g\rangle|^2\nonumber\\
    &\times\delta(E_{ZPL}^0 + \varepsilon_{n_e} - E_\lambda-\varepsilon_{n_g})
\label{eq:general_cross_section}
\end{align}

where $C^\prime$ contains physical constants, $|\langle n_e|n_g\rangle|^2$ is the overlap integral between the initial and final vibrational states, $\mathcal{B}(n_g,T)$ is the thermal occupation probability of the initial vibrational state at temperature T, $\varepsilon_{n_g}$ and $\varepsilon_{n_e}$ are the energies of the ground and excited vibrational states respectively, the delta function selects the elements of the double sum which are resonant with the excitation, and $|\hat{\mu}|^2=|\langle^2E|\hat{\mu}|^2A_2\rangle|^2$ is the optical dipole matrix element between the electronic states \cite{lax52, razinkovas21}. The electronic and nuclear components can be treated separately by making the crude adiabatic approximation, in which the vastly different time scales of the electronic and nuclear evolution decouple their relative dynamics \cite{mishra22, bersuker12}.

\subsection{Quasi-Continuum model}\label{sec:quasimodel}

\subsubsection{Model derivation}

\begin{figure}
    \centering
    \includegraphics[width=0.999\linewidth]{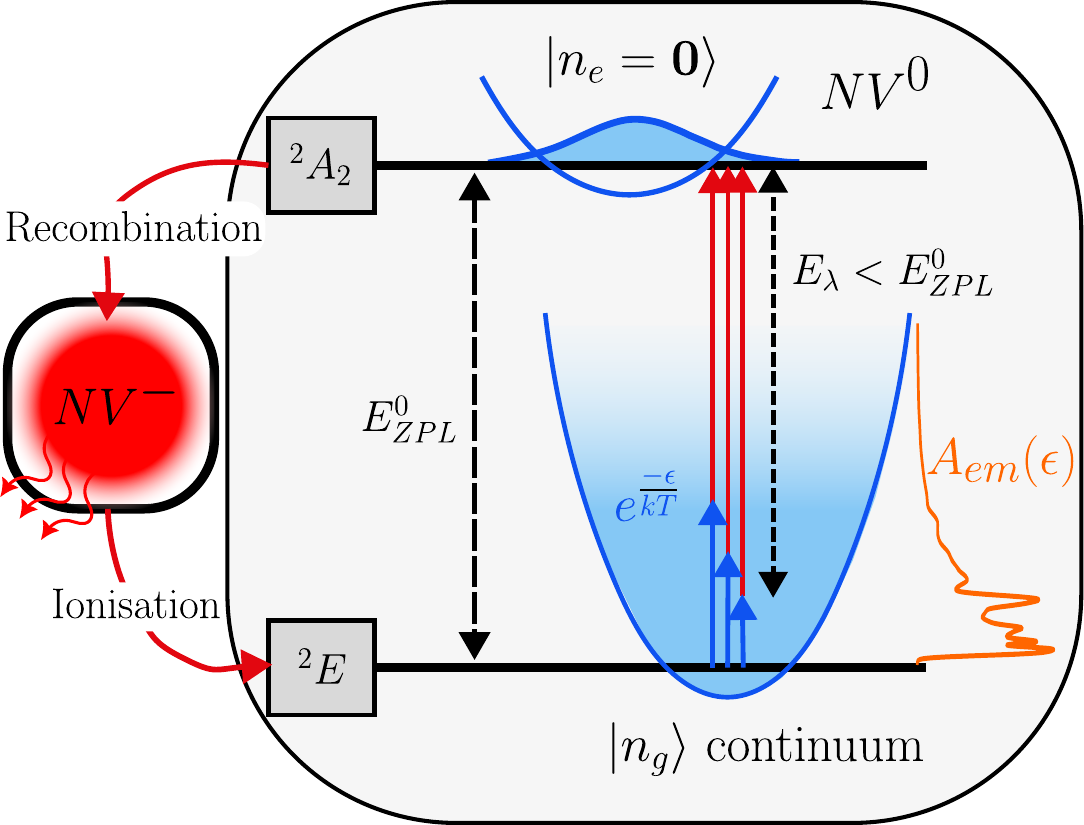}
    \caption{Schematic of the quasi-continuum model. On the right, the NV$^0$ transition is achieved by the combination of a phonon with energy $\varepsilon$, (blue arrow) and a photon (red arrow). The transition rate is found by summing over the phonons with the energy required to drive the transition, $\varepsilon\geq E_{ZPL}^0-E_\lambda$. The spectral density function, $A_{em}(\varepsilon)$, represents the strength of the coupling of phonons with energy $\varepsilon$ to the $\ket{n_e=\mathbf{0}}$ excited vibrational state. From the excited state of NV$^0$, the system can be recombined to NV$^-$. In NV$^-$, the system is directly excited by the incident photons. This leads to a burst of photon emissions leading to a bright state which can be observed experimentally. Eventually, the system is ionised and returned back to the optically dark NV$^0$ state, restarting the cycle.}
    \label{fig:quasicont}
\end{figure}

In order to obtain a minimal, analytically tractable description of the anti-Stokes excitation rate, we introduce a quasi-continuum model based on two approximations. These approximations reduce the full vibronic sum in Eq. (\ref{eq:general_cross_section}) to a form directly related to the commonly observed photon emission spectrum of NV$^0$, while retaining the essential temperature and detuning dependence of the process.  A schematic of this model is shown in Fig. \ref{fig:quasicont}.

Firstly, we approximate the ground vibrational state space with a quasi-continuous manifold of modes parametrised by a single energy variable $\varepsilon$. This is justified, especially at higher temperatures, because the electronic ground state couples to a large number of phononic modes with a large combination of vibrational configurations. This leads to a dense set of states which fill the low-energy vibrational state space quasi-continuously and which are scaled by the Boltzmann factor $e^{-\varepsilon/k_BT}$.

Secondly, we restrict the final state of the transition to the be the $\ket{n_e=\mathbf{0}}$ vibrational ground state in the excited electronic manifold. This is a reasonable simplification as transitions terminating in the lowest excited vibrational level minimize the required initial vibrational energy and therefore are not as strongly suppressed by the Boltzmann factor. As a result the dominant terms of the sum over the states are minimally affected by the approximation. This approximation is expected to perform better at lower temperatures where low energy vibrational states will dominate the dynamics.

Using these two assumptions we can express Eq. (\ref{eq:general_rate}) as

\begin{align}
    R(\lambda, T) &= CP_\lambda|\hat{\mu}|^2\sum_{\varepsilon=E_{ZPL}^0-E_\lambda}^{E_{ZPL}^0}\mathrm{e}^{-\varepsilon/k_BT}\times\nonumber\\
    &\sum_{n_g}|\langle  n_e=\mathbf{0}|n_g\rangle|^2\delta(\varepsilon +\varepsilon_{n_e=\mathbf{0}}-\varepsilon_{n_g})  
\label{eq:approx2rate}
\end{align}

where the sum over the final states has become a sum over all the phonon energies, $\varepsilon\geq E_{ZPL}^0-E_\lambda$, which have enough energy to drive the transition. This change in summation variables is necessary to sum over all possible states while retaining the assumption that the final state is $\ket{n_e=\mathbf{0}}$. The delta function selects the elements of the ground state vibrational quasi-continuum which are resonant with the transition to the $\ket{n_e=\mathbf{0}}$ excited state.

The strength of this approach is that in Eq. (\ref{eq:approx2rate}) we can recognise a well studied quantity, the spectral density function of photon emission \cite{razinkovas21, thiering2024}:

\begin{equation}
    A_{em}(\varepsilon)=\sum_{n_g}|\langle n_g|n_e=\mathbf{0}\rangle|^2\delta(\varepsilon +\varepsilon_{n_e=\mathbf{0}}-\varepsilon_{n_g})  
\end{equation}
\begin{figure*}
\centering
    \subfloat[\label{fig:approx_spec}]{%
        \includegraphics[width=0.45\textwidth]{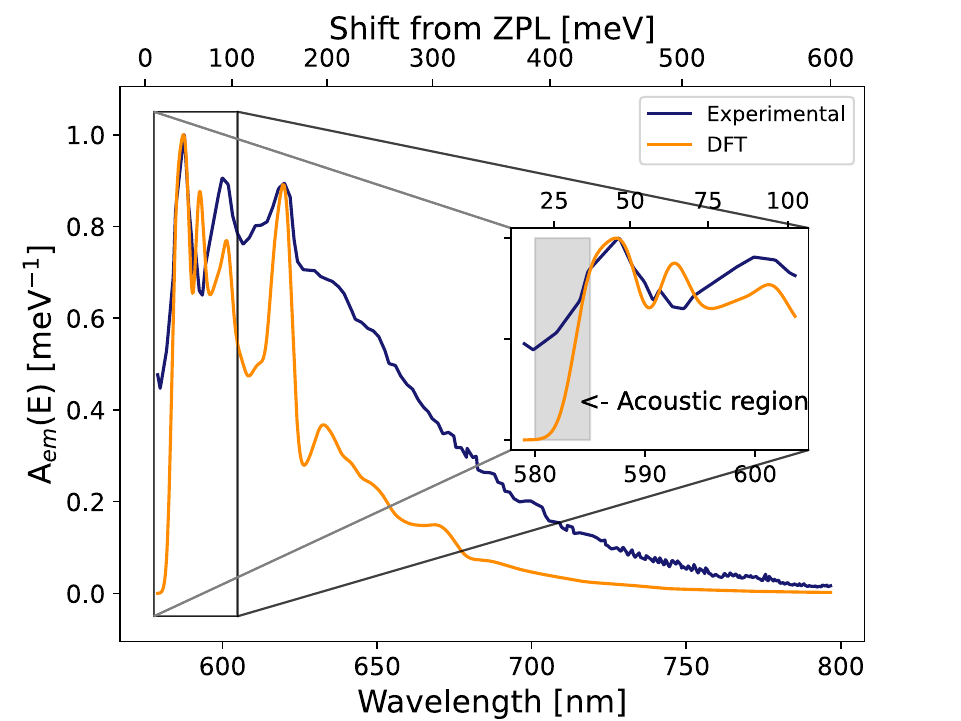}}
    \subfloat[\label{fig:approx_fit}]{%
        \includegraphics[width=0.45\textwidth]{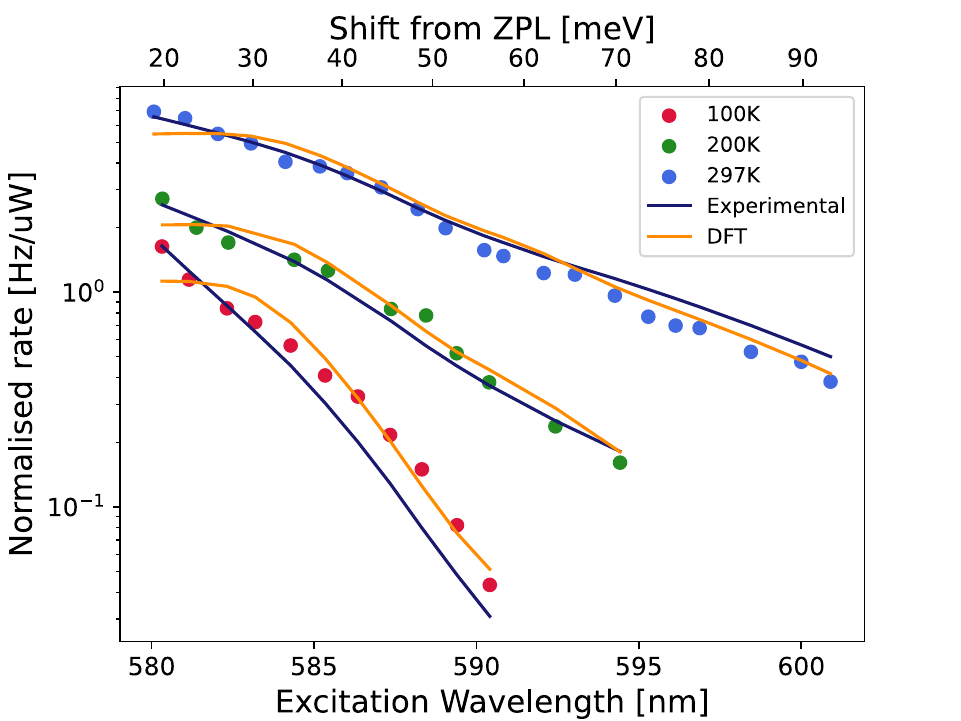}}
    \caption{(a) The experimentally measured (blue) and DFT derived (orange) emission spectra. The data was replotted from \cite{thiering2024}. The inset shows the low energy acoustic phonon regime. (b) The transition rates calculated from the quasi-continuum model fit to experimental data. The blue fit lines are generated using the experimental emission spectrum and the orange lines are generated from the DFT spectrum in panel (a). Data is displayed on a log scale.}   
    \label{fig:approxmodel}
\end{figure*}

where $\varepsilon$ is the detuning of emitted photons from the ZPL. The emission spectral function of NV$^0$ has been measured experimentally and has also been theoretically calculated using density functional theory (DFT) \cite{thiering2024}. With it, we can express the power density normalised transition rate as

\begin{align}
    \frac{R(\lambda, T)}{P_\lambda} &= C|\hat{\mu}|^2\sum_{\varepsilon=E_{ZPL}^0-E_\lambda}^{E_{ZPL}^0}\mathrm{e}^{-\varepsilon/k_BT}A_{em}(\varepsilon)
\label{eq:quasicont_eq}
\end{align}

To our knowledge, the emission spectral function has not previously been identified explicitly as the kernel governing anti-Stokes excitation rates in this regime. This equation is relatively simple to implement, however, it does require approximations as well as existing measurements of the emission spectrum. 

\subsubsection{Quasi-continuum model results}

The quasi-continuum model was fit to the transition rate data using an experimentally measured emission spectrum and a spectrum which was calculated using DFT \cite{thiering2024}. The emission spectra are shown in Fig. \ref{fig:approxmodel}(a) and the results of the fits are shown in Fig. \ref{fig:approxmodel}(b).

The DFT derived model deviates from the experimental data at wavelengths closer to the ZPL. This is likely caused by the difficulty in resolving low energy phonons in the DFT formalism due to finite-size effects \cite{razinkovas21}. Figure \ref{fig:approxmodel}(a) shows the acoustic region of the emission spectra. The DFT spectrum drops to zero in this region compared with the experimentally observed spectrum. This suggests that acoustic phonons play a key role in anti-Stokes shifted excitation at wavelengths close to the ZPL.

Conversely, the model derived from the experimental spectrum (blue lines) reproduces the measured rates more accurately, especially at elevated temperatures. In the high temperature regime, many phonon modes are thermally populated and the excitation process involves a large number of weakly contributing modes. Consequently, the detailed structure of individual modes becomes less important, and the overall spectral density of phonon-assisted transitions combined with the Boltzmann distribution governs the transition rate.

However, at low temperatures the model derived from the experimental spectrum deviates from the measured rates. In this regime, anti-Stokes excitation is likely dominated by a small subset of phononic modes, which may be obscured by the large number of modes integrated into the experimental spectrum of emission.  

Overall, this model accurately captures the qualitative features of the transition rate and provides a simple means to calculate the transition rate at higher temperatures. This simplicity, along with the accuracy of the model at high temperatures, makes it a useful tool for calculations of the expected anti-Stokes shifted transition rate in other defect systems for which the emission spectrum of the defect is known. The model also hints at the importance of low-energy acoustic modes, however, it fails to completely capture the vibrational modes which dominate the low temperature rates. As a result, we develop a second model which explicitly treats individual modes of vibration. This aims to more accurately understand the precise vibrational structure which contributes to anti-Stokes shifted excitation in NV$^0$. 

\subsection{Effective phonon mode model}\label{sec:effectivemodel}

\subsubsection{Model derivation}

\begin{figure}
    \centering
    \includegraphics[width=0.99\linewidth]{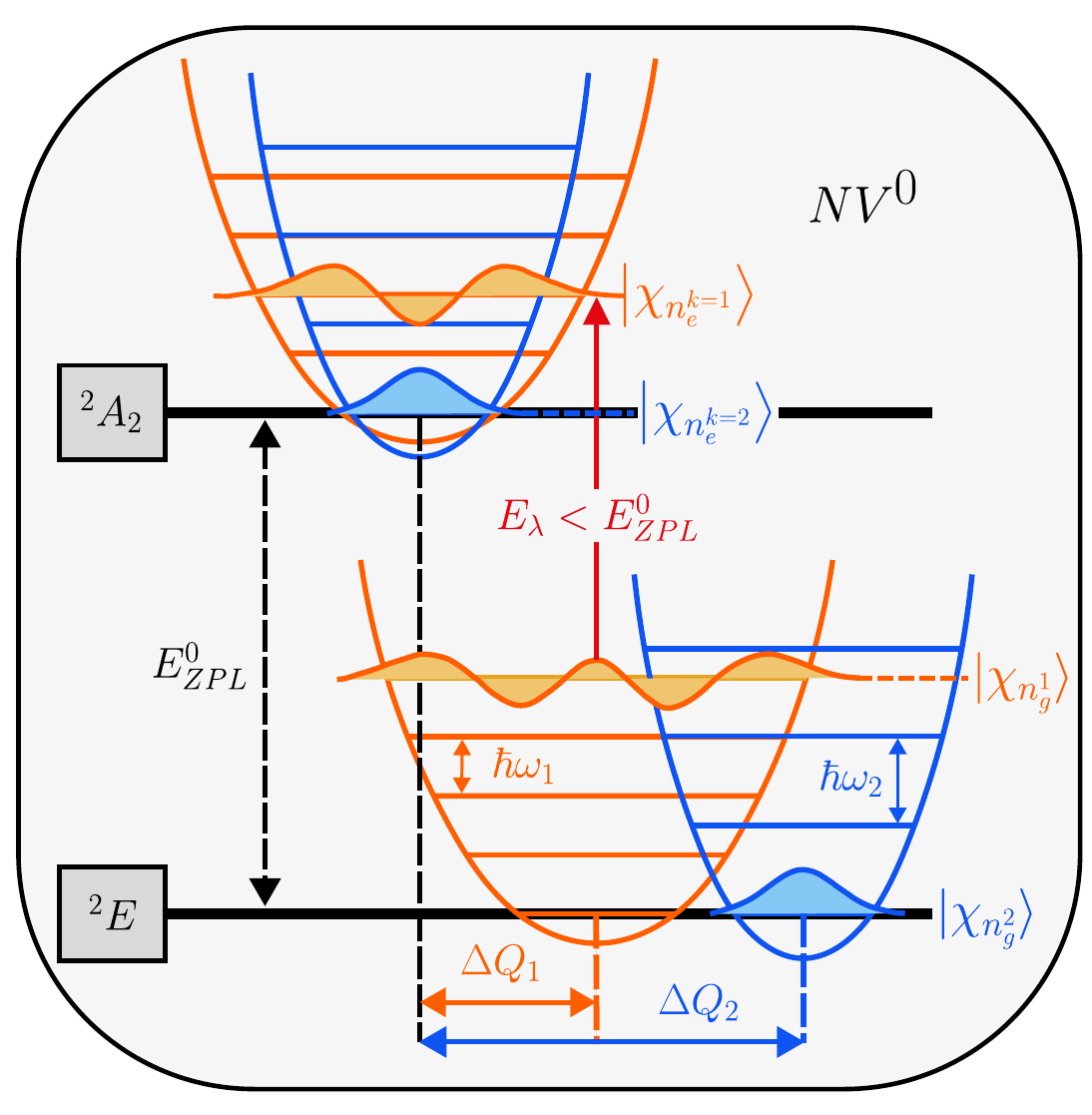}
    \caption{Configuration coordinate diagram illustrating the effective mode model. Two representative modes, $k=1,2$, are shown as harmonic oscillators with eigenstates $|\chi_{n_g^k}\rangle$ and $|\chi_{n_e^k}\rangle$ displaced by $\Delta Q_k$. Sub-ZPL ($E_\lambda<E_{ZPL}^0$) excitation occurs via absorption from thermally populated vibrational states, enabled by Frank-Condon overlaps between the displaced oscillators.}
    \label{fig:effective_schematic}
\end{figure}

In principle, the diamond lattice contains a very large number of vibrational normal modes. Within the harmonic approximation, these modes may be treated as independent quantum harmonic oscillators (see Section IV of the SM \cite{SuppMat} for a detailed derivation and references \cite{alkauskas2014, Kresse95, razinkovas21, mcquarrie2000} therein). A scematic of this transition is shown in Fig. \ref{fig:effective_schematic}. The states of the vibrational modes in the ground and excited electronic states can be expressed as products of harmonic oscillator wavefunctions

\begin{align*}
    \ket{n_g}&=\prod_{k=1}^{N}|\chi_{n_g^k}\rangle(Q_k-\Delta Q_k)\\
    \ket{n_e}&=\prod_{k=1}^{N}|\chi_{n_e^k}\rangle(Q_k)
\end{align*}

where $|\chi_{n_g^k}\rangle$ and $|\chi_{n_e^k}\rangle$ are the harmonic oscillator wavefunctions and $n_g^k$, $n_e^k$ are the number of vibrational quanta currently occupied in the $k$-th oscillator \cite{davies81}. The equilibrium configuration of the lattice differs between the ground and excited electronic states, leading to displaced harmonic potentials along each vibrational coordinate, $Q_k$. The equilibrium position of each oscillator is therefore displaced by a small distance $\Delta Q_k$ \cite{davies81, razinkovas21}. 

The strength of electron-phonon coupling for each mode is quantified by the partial Huang-Rhys factor

\begin{equation}
    S_k=\frac{\omega_k(\Delta Q_k)^2}{2\hbar}
\end{equation}

where $\omega_k$ is the fundamental frequency of the mode. This factor represents the average number of phonons which are excited in this mode during a transition \cite{davies81, razinkovas21}. The overlap between the displaced harmonic oscillator wavefunctions can be expressed analytically in terms of $S_k$ as \cite{moller96, ianchello98}

\begin{multline}\label{eq:overlap_main}
    |\langle\chi_{n_e^k}|\chi_{n_g^k}\rangle|^2 =\mathrm{e}^{-S_k}\,(n_g^k!\,n_e^k!)\\
    \times\left(\sum_{\ell=0}^{\mathrm{min}(n_g^k, n_e^k)}(-1)^\ell\frac{S_k^{\frac{1}{2}(n_g^k+n_e^k)-\ell}}{\ell!(n_g^k-\ell)!(n_e^k-\ell)!}\right)^2
\end{multline}

While the Franck–Condon formalism permits excited initial and final vibrational states, this has not been implemented in previous quantitative modelling of absorption and emission in NV centres \cite{lax52, razinkovas21}. It is commonly assumed that the system will occupy the ground vibrational state before absorption or emission, however, this assumption is not valid in the anti-Stokes regime. Our model thus utilizes the formalism more generally than previous studies of the NV. 

The occupation probability of the ground state vibrational level, $\ket{n_g}$, is found using Boltzmann statistics to be \cite{mcquarrie2000} 

\begin{equation}\label{eq:boltzmann}
    \mathcal{B}(n_g,T) \propto \exp\left[-\sum_k\frac{n_g^k\hbar\omega_k}{k_BT}\right]
\end{equation}

Summing over many modes is computationally expensive. We therefore adopt an \textit{effective mode model} in which the many lattice modes are represented by a small number of oscillators. In this model, the effective modes are Lorentzian broadened because they stand in for a large number of modes. The Lorentzian lineshape is given by

\begin{equation}\label{eq:lorentzian}
    \mathcal{L}(E, \Gamma) = \frac{1}{1+\left(\frac{E}{\Gamma/2}\right)^2}
\end{equation}

where $\Gamma$ is the full-width-half-maximum of the broadening.

In this effective mode regime, the power density normalised rate of the transition can be written as

\begin{multline}
    \frac{R(\lambda, T)}{P_\lambda}\approx C^\prime |\hat{\mu}|^2\sum_{\{n_g\}}\sum_{\{n_e\}}\mathrm{exp}\left[-\sum_k\frac{n_g^k\hbar\omega_k}{k_BT}\right]\times\\
    \mathcal{L}(E_{ZPL}^0 + \sum_k\hbar\omega_k(n_e^k-n_g^k)-E_\lambda, \Gamma)\left(\prod_k|\langle\chi_{n_e^k}|\chi_{n_g^k}\rangle|^2\right)
    \label{eq:effrate}
\end{multline}

where $C^\prime$ is a constant and $k$ runs only over the effective modes discussed above. This equation sums over all possible combinations of the initial and final occupations of the effective phonon modes. It computes the Franck-Condon overlap integral between the harmonic oscillator wavefunctions using Eq. (\ref{eq:overlap_main}). This is scaled by the Boltzmann occupation probability of the initial vibrational state (Eq. (\ref{eq:boltzmann})) and by the Lorentzian broadened resonance condition of the excitation laser. 

%The equilibrium position of a nucleus, $m$, in the electronic ground state, $\mathbf{R}_m^g$, is displaced from its equilibrium position in the electronic excited state, $\mathbf{R}_m^e$, as the different electronic wavefunctions impart different forces on the nucleus. The displacement of the nucleus under a change in the electronic state is given by

%\begin{equation*}
%    \Delta\mathbf{R}_m=\mathbf{R}_m^e-\mathbf{R}_m^g
%\end{equation*}

%which decays as we consider nuclei located further from the NV centre. Nuclei at a distance greater than 3 lattice constants from the NV experience almost no displacement due to the highly localised electronic wavefunctions and the stiffness of the diamond lattice \cite{alkauskas2014}. 

\begin{figure*}
\centering
    \subfloat[\label{fig:analytic_spec}]{%
        \includegraphics[width=0.45\textwidth]{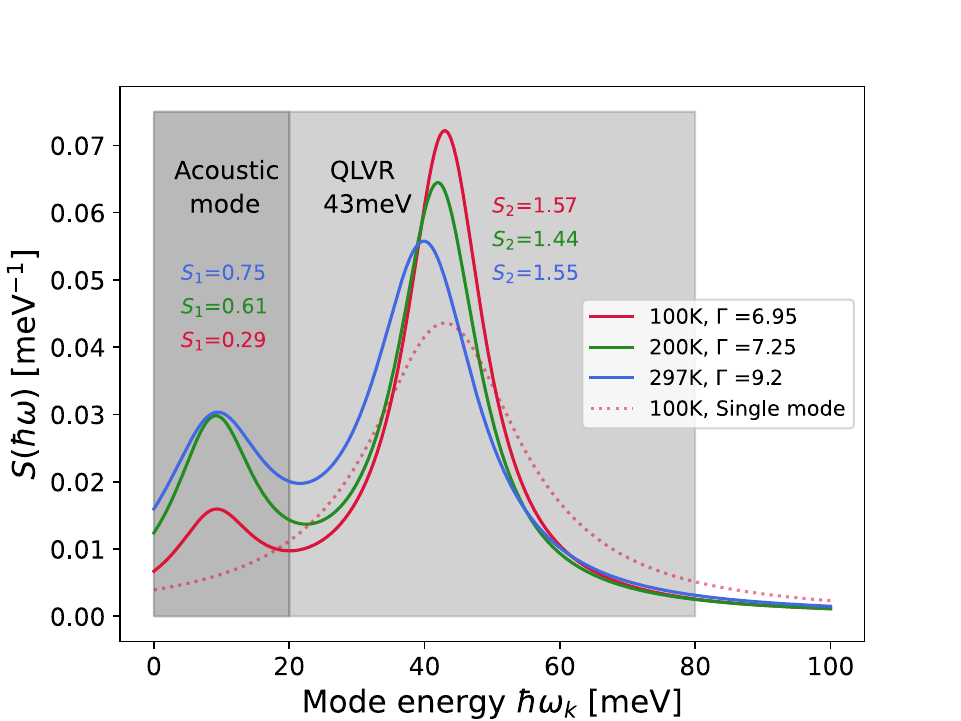}}
    \subfloat[\label{fig:analytic_fits}]{%
        \includegraphics[width=0.45\textwidth]{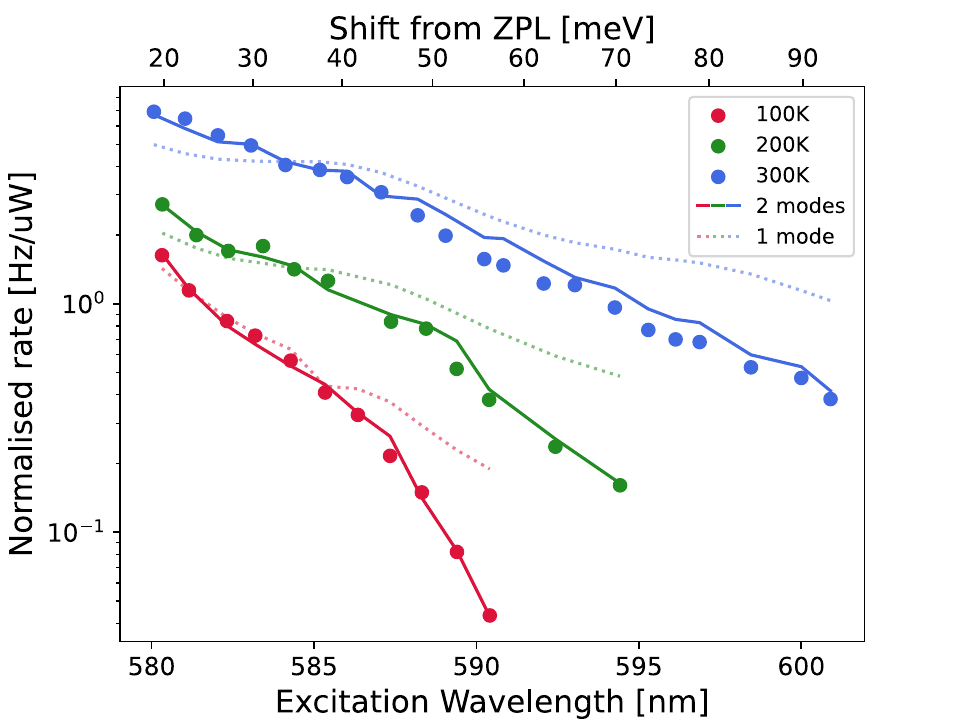}}
    \caption{(a) The spectrum of electron-phonon coupling generated from the effective vibrational modes found from the fitting procedure. The dashed line shows the single mode fit for the 100\,K data. The solid lines are the 2 mode fits for which the best fitting partial HR factors $\{S_1, S_2\}$ and the Lorentzian broadening terms $\Gamma$[meV] are shown. (b) Experimentally measured transition rates at 100\,K, 200\,K and 300\,K, fit using the effective phonon mode model.}
    \label{fig:analyticmodel}
\end{figure*}

\subsubsection{Fitting procedure}

The model was fit to the experimental data. Firstly, a single effective mode was fit using a mode energy at the experimentally observed $\hbar\omega_k=$43\,meV quasi-local vibrational resonance (QLVR) \cite{thiering2024, Su19}. This mode of vibration contributes strongly to the emission spectrum of NV$^0$ and involves lattice ions close to the NV centre \cite{Su19}. Secondly, another mode was added at 9\,meV in addition to the 43\,meV mode. This mode was added in order to represent a bath of low energy acoustic phonon modes which were found to be crucial in the quasi-continuum model to achieve quantitative agreement. Adding a third mode was found not to improve the fit quality significantly and was highly computationally taxing.

For each mode, the partial Huang-Rhys factor, $S_k$, was used as a fitting parameter. The broadening term, $\Gamma$, was shared by each mode and was also used as a fitting parameter. The resulting electron-phonon coupling spectrum, $S(\hbar\omega)$, is found by summing over the Lorentzian broadened modes. It is given by

\begin{equation}
    S(\hbar\omega) = \sum_k\mathcal{L}(\hbar\omega_k-\hbar\omega,\Gamma)\cdot S_k
\end{equation}

and is shown in Fig. \ref{fig:analyticmodel}(a).

\subsubsection{Effective mode model results}

The fit results are shown in Fig. \ref{fig:analyticmodel}(b). The single 43\,meV mode model was unable to accurately represent the data, and the addition of the 9\,meV acoustic mode was required to achieve quantitative agreement. 

The necessity of the acoustic mode further indicates that, for small detunings from the ZPL energy, anti-Stokes excitation is facilitated by the availability of low-energy phonons capable of bridging small energy deficits between the excitation photon and the electronic transition. The fitted partial Huang–Rhys factors associated with the acoustic effective mode, shown in Fig. \ref{fig:analyticmodel}(a), were observed to increase with temperature. This temperature dependence should not be interpreted as a change in microscopic coupling strength, which is dependent only on the geometry of the phonon modes, but rather as reflecting the increasing participation of thermally populated acoustic modes \cite{alkauskas2014}.

The 43\,meV mode contributes primarily at larger excitation wavelength detunings and is especially prominent in the low temperature data. At 100\,K, the transition rate bulges at detunings around 43\,meV and then drops off rapidly. This suggests that at low temperatures, the 43\,meV mode acts as an upper bound for phononic energy and thus the transition rates at detunings greater than this energy will be strongly suppressed. As the temperature increases, this mode is mixed with the highly occupied acoustic modes and the bulge in the transition rate at 43\,meV is lost within the statistical ensemble of vibrational states in the 300\,K data.

The success of this two-mode description indicates that the vibrational environment of NV$^0$ is well described by the interplay between a localised vibrational resonance and a low-energy acoustic bath.

The Lorentzian broadening parameters extracted from the fits are comparable to those expected from finite phonon lifetimes in the fs range. Further, the broadening increases with temperature, reflecting the shorter lifetimes of the phonon states expected at higher temperature. This behaviour is consistent with a phonon-mediated transition.

\subsection{Comparison of modelling approaches}

The quasi-continuum model performs better at higher temperatures where the large number of occupied phonon modes are represented in the emission spectrum function. In contrast, the effective-mode model captures the low temperature dependence of the transition rate more accurately. In the low temperature range, the structure of the individual modes is prominent and is well modelled by the effective mode approach. The crossover between these temperature regimes demonstrates the changing behaviour of anti-Stokes shifted excitation, shifting from depending on a small number of dominant modes to a statistical ensemble of vibrational states.

\section{Conclusion}
In this work, we have identified the microscopic origin of anti-Stokes shifted charge cycling in the nitrogen-vacancy centre by combining single-defect measurements with quantitative vibronic modelling. Measurements of a spatially isolated NV centre reveal that the transition from NV$^0$ to NV$^-$ under sub-resonant excitation is strongly suppressed at cryogenic temperatures and at large detuning of the excitation photon energy below the NV$^0$  ZPL wavelength. This is highly suggestive that a phonon-assisted mechanism underlies the observed behaviour.

The measurements further reveal how specific components of the vibrational environment contribute to this process. Two complementary modelling approaches show that anti-Stokes excitation arises from a interplay between different phonon modes. Near the zero-phonon line, thermally occupied low-energy acoustic phonons dominate the excitation dynamics by bridging small energy deficits between the optical excitation and the electronic transition. At larger detunings, excitation is assisted by a quasi-local vibrational resonance at 43\,meV. The success of a minimal effective-mode description demonstrates that the essential physics of the transition can be captured by an acoustic phonon bath combined with a quasi-localized vibrational mode near 43meV.

These results provide a quantitative methodology for predicting charge-conversion rates under sub-resonant excitation conditions. Because sub-ZPL excitation schemes are increasingly used for spin-to-charge conversion and non-destructive readout protocols, identifying the phonon-assisted mechanism underlying charge cycling enables more systematic optimisation of initialization fidelity and charge stability in NV-based quantum sensors.

\begin{acknowledgments}
We would like to thank Gerg\H{o} Thiering and Philipp Vetter for the fruitful discussions.  
We thank the Ulm Center for Nanotechnology and Quantum Material for the metal deposition. We thank Professor Dr. Christine Kranz, Dr. Gregor Neusser, and the Focused Ion Beam Center at Ulm University for their support with FIB milling.

This work was supported by the QuantumBW innovation initiative and the Baden-W\"{u}rttemberg Ministry of Science, Research and the Arts via the IQST Graduate School @QuantumBW, the BMFTR via projects QSENS (03ZK110AB), QR.X (16KISQ006) and Diaqnos (13N16463), the Carl Zeiss Foundation via the Ultrasens-Vir project P2022-06-007, the DFG via projects 387073854, 445243414, 491245864, 546850640, and 560722984, the joint DFG/JST ASPIRE program via project 554644981, the ERC via the Synergy grant HyperQ via project 856432, and the EU via the H2020 projects FLORIN (101086142) and SPINUS (101135699).
\end{acknowledgments}

\bibliography{bibliography}
\end{document}

\title{Supplemental Material:\\
``Phonon-assisted charge-cycling of nitrogen-vacancy centres in diamond''}

\author{Michael Olney-Fraser}%
\affiliation{Institute for Quantum Optics, Ulm Univesity, Albert-Einstein-Allee 11, 89081 Ulm, Germany}
\affiliation{Integrated Quantum Science and Technology (IQST), Ulm University, Albert-Einstein-Allee 11, Ulm 89081, Germany}

\author{Stefan Dietel}
\affiliation{Institute for Quantum Optics, Ulm Univesity, Albert-Einstein-Allee 11, 89081 Ulm, Germany}

\author{Jens Fuhrmann}
\affiliation{Institute for Quantum Optics, Ulm Univesity, Albert-Einstein-Allee 11, 89081 Ulm, Germany}

\author{Lev Kazak}
\affiliation{Institute for Quantum Optics, Ulm Univesity, Albert-Einstein-Allee 11, 89081 Ulm, Germany}

\author{Fedor Jelezko}%
\affiliation{Institute for Quantum Optics, Ulm Univesity, Albert-Einstein-Allee 11, 89081 Ulm, Germany}
\affiliation{Integrated Quantum Science and Technology (IQST), Ulm University, Albert-Einstein-Allee 11, Ulm 89081, Germany}

\maketitle

\section{Experimental setup and characterisation of NV centre}

\begin{figure}
    \centering
    \includegraphics[width=0.99\linewidth]{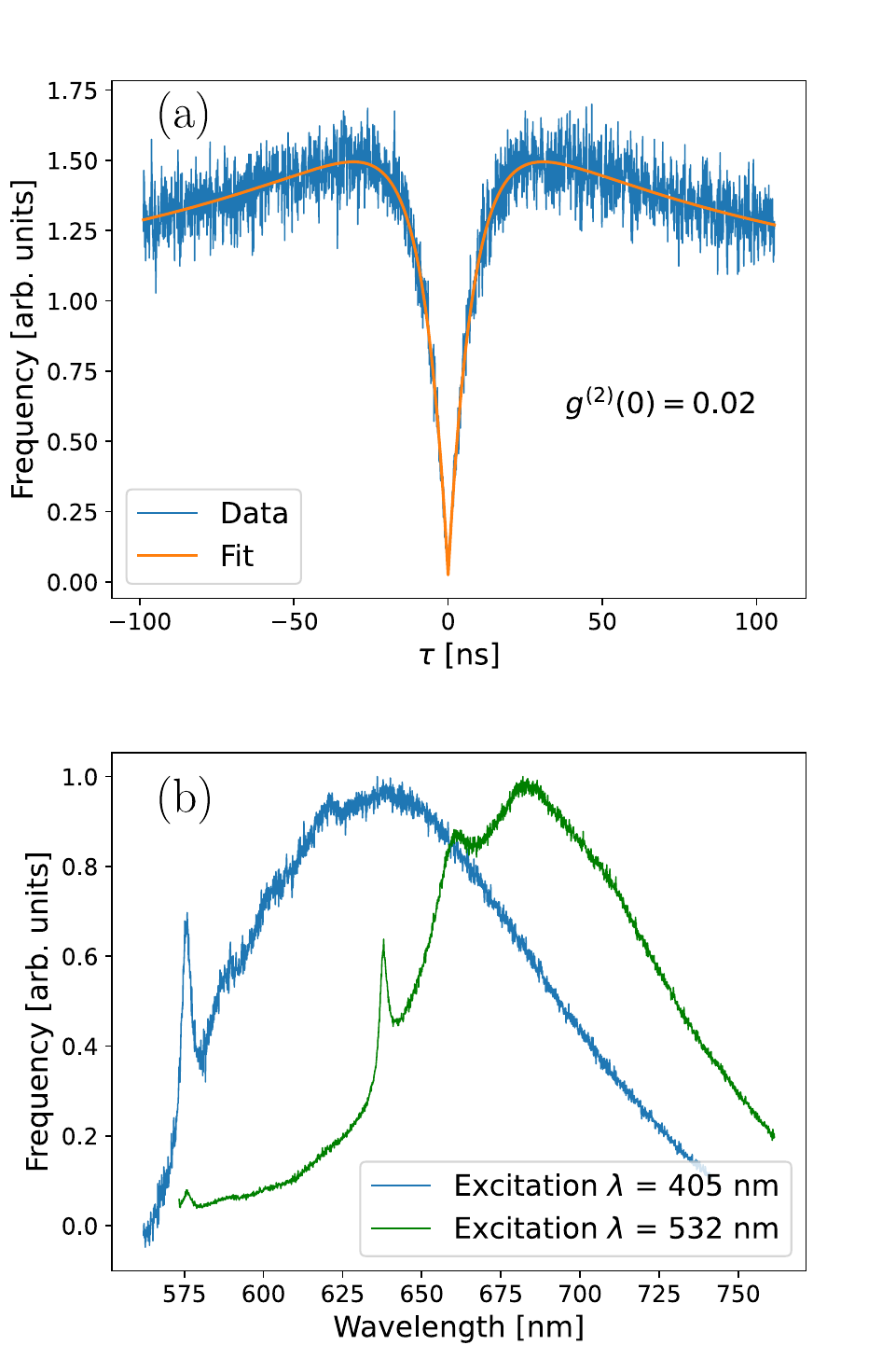}
    \caption{(a) Autocorrelation function measurement on the NV used in this experiment. (b) NV emission spectrum under 532nm (green) and 405nm (blue) excitation.} 
    \label{fig:sample}
\end{figure}

A custom-built confocal microscope with 532\,nm and 405\,nm excitation lasers was used to locate NV centres which were activated at favourable positions within the SILs. The observed photoluminescence signal of the NV used in this study was enhanced by a factor of 6 by the SIL. To verify that the emissions originated from a single NV, the $g^{(2)}(\tau)$ autocorrelation function was measured by using a Hanbury-Brown-Twiss interferometer with a FastCommTech fast counting card to correlate the signals from the two APDs \cite{brown56, zwiller01}. The $g^{(2)}(\tau)$ measurement is shown in Figure \ref{fig:sample}a and confirms that the signal comes from a single quantum emitter. Furthermore, to ensure that the emitter was in fact an NV centre, optical spectra were measured under 532\,nm and 405\,nm excitation using a Princeton Instruments Spectrometer. The spectra are shown in Figure \ref{fig:sample}b and clearly show the NV$^-$ and NV$^0$ zero-phonon lines and characteristic phonon sidebands.   

\section{4\,K data}

\begin{figure}[b]
    \centering
    \includegraphics[width=0.95\linewidth]{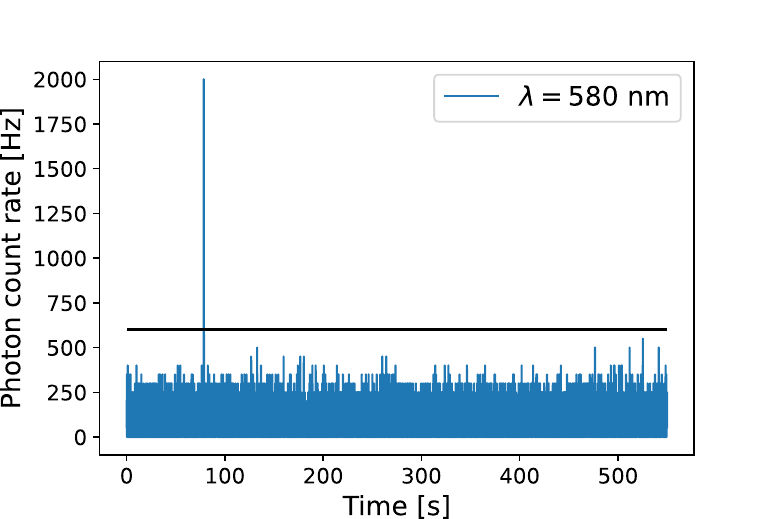}
    \caption{Time series photon count rate data taken at 4\,K under excitation with a 580\,nm laser. A single bright state transition is observed above the threshold.}
    \label{fig:4K}
\end{figure}

At 4\,K, the transition rate from NV$^0$ to NV$^-$ was too slow for a statistically robust measurement of the rate to be made. Figure \ref{fig:4K} shows an example time series measurement taken with an excitation wavelength of 580\,nm and a laser power of 30\,$\mu$W. This power is significantly higher than those used to perform the higher temperature measurements shown in the main text, and the wavelength is close to the ZPL of NV$^0$. As can be seen, only a single bright state was observed over the 600\,s time series measured. This indicates that, while transitions still occur extremely rarely, they are all but completely eliminated at 4\,K. As a result, we use the 4\,K measurement as a phenomenological observation that the transition is quenched at low temperatures without pursing a quantitative analysis. 

\section{Mean first passage analysis of the transition rate}

\begin{figure}
    \centering
    \includegraphics[width=0.95\linewidth]{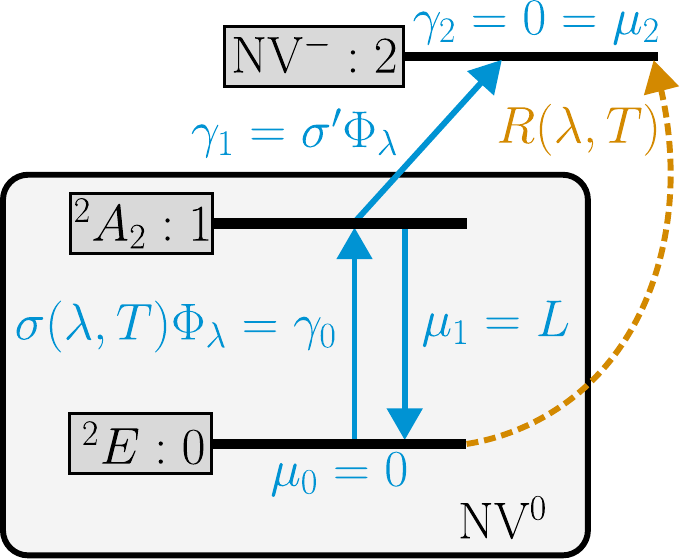}
    \caption{The NV$^0$ ground state (state 0), excited state (state 1) and the NV$^-$ charge state (state 2) are shown. Connecting them are the transition at the rates between the states. The Markov-chain rates $\lambda_i$ and $\mu_i$ are shown as well as the photophysical transition rates. The transition rate from state 0 to state 2, $R(\lambda,T)$, is the rate measured and modelled in the main text.}
    \label{fig:mfp}
\end{figure}

The average transition rate from the ground state of NV$^0$ back to NV$^-$ can be found by treating the system as a continuous time Markov chain and calculating the mean first passage time \cite{allen10}. In the continuous Markov chain formalism, we define the state indices 0 (ground state of NV$^0$), 1 (excited state of NV$^0$) and 2 (NV$^-$). We also define the excitation rates, $\gamma_i$, and relaxation rates, $\mu_i$, leaving the state $i$. A schematic of the Markov chain is shown in Figure \ref{fig:mfp}.

The expected value of the time taken to transition from state $i$ to state $i+1$ can be defined as \cite{allen10}

\begin{equation}
\mathbb{E}(T_{i,i+1})=\frac{1}{\gamma_i}+\frac{\mu_i}{\gamma_i}\mathbb{E}(T_{i-1,i})
\end{equation}

We are interested in the average time taken to transition from state 0 to state 2, which can be represented using the rates shown in Figure  \ref{fig:mfp} as

\begin{align}
\mathbb{E}(T_{0,2}) &= \mathbb{E}(T_{0,1})+\mathbb{E}(T_{1,2})\\
&=\left(\frac{1}{\gamma_0}\right)+\left(\frac{1}{\gamma_1} + \frac{\mu_1}{\gamma_1}\frac{1}{\gamma_0}\right)\\
&=\frac{\gamma_0+\gamma_1+\mu_1}{\gamma_0\gamma_1}
\end{align}

Inverting this expression, we can find the effective transition rate from the ground state of NV$^0$ to NV$^-$

\begin{equation}\label{eq:appendix_effrate}
R = \frac{\gamma_0\gamma_1}{\gamma_0+\gamma_1+\mu_1}
\end{equation}

This rate can be expressed in terms of the photophysical transition rates between each of the states. Under laser excitation with wavelength $\lambda$, we define the flux of incident photons from the excitation laser, $\Phi_\lambda$. Next we define the cross section of absorption from state 0 to state 1, $\sigma(\lambda,T)$, where $T$ is the temperature. In the case of anti-Stokes shifted excitation, this cross section depends on the occupation of phononic states as modelled in the main text. We next define the cross section of the excitation transition from state 1 to 2, $\sigma^\prime$, and the spontaneous relaxation rate from state 1 to 0, $L$. With these photophysical values we can define the excitation and decay rates as:

\begin{align}
\gamma_0&=\sigma(\lambda, T)\Phi_\lambda\\
\gamma_1&=\sigma^\prime\Phi_\lambda\\
\mu_1&=L
\end{align}

Inserting these into equation \ref{eq:appendix_effrate}, and rearranging we can write the transition rate as

\begin{equation}
R(\lambda, T) = \sigma(\lambda,T)\Phi_\lambda\left( \frac{1}{\frac{\sigma(\lambda,T)}{\sigma^\prime}+1+\frac{L}{\sigma^\prime\Phi_\lambda} }\right)
\end{equation}

The transition from the excited state of NV$^0$ back to NV$^-$ is driven directly by the laser, leading to a large cross section value, $\sigma^\prime$, which does not depend strongly on the excitation wavelength or temperature of the sample. Conversely, the excitation of the main transition of NV$^0$ in the anti-Stokes shifted regime relies on the occupation of phononic states, leading to a very small absorption cross section which depends on the excitation wavelength and temperature $\sigma(\lambda,T)$. The significantly different magnitudes of these cross sections allow us to neglect the ratio $\sigma(\lambda,T)/\sigma^\prime$. The transition rate can therefore be written as 

\begin{equation}\label{eq:appendix_rate}
R(\lambda,T) = \sigma(\lambda,T)\Phi_\lambda\left( \frac{1}{1+\frac{L}{\sigma^\prime\Phi_\lambda}}\right)
\end{equation}

The photon flux $\Phi_\lambda$ was held constant across the entire wavelength range studied in the experiment, which makes the term in brackets in equation \ref{eq:appendix_rate} constant. This leads to the expression

\begin{equation}
R(\lambda,T) = C\sigma(\lambda,T)\Phi_\lambda
\end{equation}

which is the equation used in the main text for the transition rate.

\section{Nuclear modes of vibration derivation}\label{sec:appendix_effectivemode}

The patterns of vibration which make up the states $\ket{n_g}$ and $\ket{n_e}$ are highly non-localised, spanning tens of thousands of nuclei \cite{alkauskas2014}. This is due to the fact that the small, localised displacements induced by a change in the electronic state are carried outward into the stiff diamond lattice on collective vibrational waves. 

For this reason, many ($>$10,000) nuclei must be considered when modelling the vibrational states, leading to a large number, $N$, of vibrational normal modes, one for each Cartesian direction, for each nucleus. 

The normal modes of vibration of the entire system of nuclei can be found by diagonalisation of the dynamical matrix, 

\begin{equation*}
    D_{\alpha\beta}(m,n) = \frac{1}{\sqrt{M_mM_n}}\frac{\partial F_{m,\alpha}}{\partial R_{n,\beta}}
\end{equation*}

where a nucleus $m$, with mass $M_m$, experiences a force in the Cartesian direction $\alpha$ given by $F_{m,\alpha}$, which changes according to a change in the position of a nucleus $n$, with mass $M_n$, in the Cartesian direction $\beta$ \cite{Kresse95}.

Linearizing Newton’s equations of motion about the ground state equilibrium leads to the eigenvalue problem

\begin{equation*}
    \sum_{n,\beta}D_{\alpha\beta}(m,n)\boldsymbol{\eta}^k_{n,\beta}=\omega_k^2\boldsymbol{\eta}^k_{m,\alpha}
\end{equation*}

where $k=1,...,N$ is the index for the normal modes, $\omega_k$ is the angular frequency of the $k$-th mode, and the dimensionless eigenvectors $\boldsymbol{\eta}^k\in\mathbb{R}^{N}$ contain the directions along which each nucleus, $m$, oscillates harmonically in this mode, $\boldsymbol{\eta}^k_m\in\mathbb{R}^3$. The normal mode vectors form an orthogonal basis of the $N$-dimensional configuration space of the nuclei. We make the equal mode approximation that the same solutions are valid in the excited electronic state \cite{razinkovas21}.

With this basis of vibrational modes and frequencies, the collective oscillation of all the nuclei can be described by $N$ quantum harmonic oscillators. The instantaneous displacement of a nuclei from its equilibrium position, $\mathbf{R}_m-\mathbf{R}_m^g$, can be projected onto the $k$-th mode vector, and the sum of these projections over all the nuclei gives the current displacement of the system along the $k$-th mass weighted configuration coordinate, $Q_k$,

\begin{equation*}
    Q_k=\sum_m\sqrt{M_m}(\mathbf{R}_m-\mathbf{R}_m^g)\cdot\boldsymbol{\eta}_m^k
\end{equation*}

Upon quantisation, each of these coordinates correspond to an independent quantum harmonic oscillator. The vibrational eigenstates of the system can therefore be expressed as products of single-mode harmonic oscillator states. If we let

\begin{equation*}
    n_{g(e)}=\left(n_{g(e)}^1,...,n_{g(e)}^{N}\right)\in\mathbb{N}^{N}
\end{equation*}

be a vector which denotes the number of vibrational quanta in each mode in the ground (excited) electronic state, then the ground and excited state vibrational wavefunctions can be written as the products

\begin{align*}
    \ket{n_g}&=\prod_{k=1}^{N}|\chi_{n_g^k}\rangle(Q_k-\Delta Q_k)\\
    \ket{n_e}&=\prod_{k=1}^{N}|\chi_{n_e^k}\rangle(Q_k)
\end{align*}

where $n_g^k$ and $n_e^k$ are the number of quanta in the $k$-th mode of the ground and excited states respectively and 

\begin{equation*}
    |\chi_{n}\rangle(Q) = \frac{1}{\sqrt{2^nn!}}\left(\frac{\omega}{\pi\hbar}\right)^{1/4}\mathrm{e}^{-\frac{1}{2\hbar}\omega Q^2}H_{n}\left(\sqrt{\frac{\omega}{\hbar}}Q\right)
\end{equation*}

are the harmonic oscillator wavefunctions in coordinate $Q$ with frequency $\omega$ and $n$ vibrational quanta, written in terms of the Hermite polynomials $H_n$. 

The equilibrium positions of each oscillator depend on the electronic state of the NV due to the different forces on the nuclei caused by the different electronic wavefunctions. For each nucleus, the equilibrium position in the ground electronic state is defined as $\mathbf{R}_m^g$ and in the excited state is $\mathbf{R}_m^e$. The displacement of the ground and excited state equilibria of each coordinate is given by summing over the nuclear displacements projected onto the normal modes

\begin{equation*}
    \Delta Q_k=\sum_m\sqrt{M_m}(\mathbf{R}_m^e-\mathbf{R}_m^g)\cdot\boldsymbol{\eta}_m^k=\sum_m\sqrt{M_m}\Delta\mathbf{R}_m\cdot\boldsymbol{\eta}_m^k
\end{equation*}

%These configuration coordinates behave as independent quantum harmonic oscillators. The vibrational state of the whole system is described by the number of vibrational quanta currently in each harmonic oscillator. The equilibrium positions of the configuration coordinates differ depending on the electronic state of the NV. The displacement between the ground and excited state equilibrium positions of each mode is given by  

%The independence of each mode allows the total wavefunction of the ground and excited vibrational states to be formed from the product of the wavefunctions of each mode. If we let 

The energies of the vibrational states $\ket{n_{g(e)}}$ are the sums of the harmonic oscillator level energies for each independent mode \cite{mcquarrie2000}

\begin{equation*}
\varepsilon_{n_{g(e)}}=\sum_{k=1}^N\hbar\omega_k\left(n_{g(e)}^k+\frac{1}{2}\right)
\end{equation*}

The $\ket{n_g}$ and $\ket{n_e}$ wavefunctions and their energies are used in the effective mode model in the main text. 

\bibliography{bibliography}